\documentstyle[aps,prb,preprint,epsf]{revtex}

\begin{document}
\title{Optical phonon induced frictional drag in coupled
two-dimensional electron gases} 

\author{Ben Yu-Kuang Hu}

\address{Mikroelektronik Centret,  Bygning 345\o,  
Danmarks Tekniske Universitet, DK-2800 Lyngby, Denmark}

\date{\today}
\maketitle

\begin{abstract}
The role of optical-phonons in frictional drag between 
two adjacent but electrically isolated two-dimensional electron
gases is investigated. 
Since the optical-phonons in III-V materials have a considerably larger 
coupling to electrons than acoustic phonons (which are the dominant drag
mechanism at low $T$ and large separations) it might be expected 
that the optical phonons will give a large effect high temperatures.  
The two key differences between optical and acoustic phonon mediated
drag are: (1) the optical phonon mediated interlayer interaction is 
short-ranged due to their negligible group velocity at the Brillouin 
zone center; and (2)~the typical momentum transfer for an optical 
phonon mediated scattering is relatively large.  
These considerations make optical
phonon mediated drag difficult to see in single subband GaAs systems,
but it may be possible to see the effect in double subband GaAs systems
or single subband quantum wells in a material with a lower effective
mass and lower optical phonon energy such as InSb. 
\end{abstract}
\newpage

\section{Introduction}

The study of the effects of interactions between electrons 
makes up a considerable part of current research effort in condensed matter
physics.  One system which is particularly well-suited system for the study 
of phenomena associated with electron--electron interactions is 
the two-dimensional electron gas (2DEG) in a semiconductor quantum well. 
By changing various
parameters in the system, either during the fabrication process or by the
appropriate application of gate biases, magnetic field, temperature,
etc. during the experiment, the physical conditions of the system can
be changed over a wide range of parameter space.  
Probing electrical, optical, thermal
or other properties of these 2DEGs often reveals interesting effects
due to interactions. 

Recently, it has been possible to make a direct study of interparticle 
interactions in 2DEGs, based on an proposal made some time
ago.\cite{pogr77,pric83}
Experimentalists have been able to fabricate electrically 
isolated quantum wells which are only several hundred {\AA}'s apart. 
Thus, the electrons in the adjacent wells are coupled (through Coulomb and
phonon mediated forces), but are unable to tunnel from one side 
to the other.  These remarkable structures allow the
experimentalist to probe directly the interlayer interactions
by doing a ``simple'' transport measurement; i.e., turning
on a current in one layer, and measuring the voltage built up in
the second.\cite{exper1,exper2,exper3,exper3a,exper4,exper5}  
These experiments provide a unique opportunity for a transport measurement 
to study effects of interparticle interaction.

The most obvious interlayer interaction is of course the Coulomb
coupling, which gives relatively featureless 
dependences (at zero magnetic field) as a function of temperature, 
relative density, etc.\cite{jauh93}  
Experiments, however, have shown that there are interesting pronounced 
features in drag which are associated with the collective modes in the 
system, indicating that under the right conditions drag is particularly 
sensitive to presence of these modes.  
So far, the drag effects from both acoustic
phonons and plasmons have been observed\cite{exper1,exper3,exper5} 
and studied theoretically,\cite{tso92,zhan93,flen94,swie95,flen95b,bons97} 
but to the best of our knowledge no work on 
optical phonon mediated drag has been published.

Looking first at acoustic phonons in drag, it may seem 
surprising that they can contribute so strongly to the drag transresistivity,  
given their very weak coupling to the electron gas (relative to the 
Coulomb interaction).  It turns out that strong phase-space effects at
particular energy and momentum transfers cause the acoustic phonons to 
dominate in spite of the weak coupling.\cite{bons97} 
An obvious question arises: if the weak acoustic
phonons give such a large contribution, can the optical phonons 
with a much larger coupling (witness the fact that scattering from
optical phonons produces a much larger resistivity than scattering off
their acoustic counterparts) give a correspondingly giant drag rate when
they come in at higher temperatures?

The answer depends on the type of system one is studying.  
There are two significant differences between
acoustic and optical phonons.  The first is that, unlike their acoustic
counterparts, the optical phonons around the Brillouin zone center
(those which are relevant for drag) have a very small group velocity.
Thus, the range of the optical phonons is much shorter than the acoustic
phonons.   In fact the longitudinal polar optical phonon mediated 
interaction has spatially exactly the same short-ranged form as the
out-of-plane Coulomb interaction in two-dimensional structures; {\sl
i.e.}, it is proportional to $\exp(-q d)$ where $\hbar q$ is the 
electron momentum transferred and $d$ is the center-to-center 
interlayer separation.
Furthermore, every optical phonon mediated scattering event 
involves a transfer of the energy of an optical phonon.
Since this energy is relatively large, conservation
of energy dictates that the momentum transfer $\hbar q$ must also be 
large, which together with the $\exp(-q d)$ form of the interaction 
implies that the optical phonon mediated drag effect is usually small. 
However, as shown below, under certain experimental circumstances
the effect may be relatively large and clearly observable.
Provided the substantial experimental hurdles can be surmounted, 
it may be possible to perform a room temperature drag experiment.

The paper is organized as follows.  In Section \ref{sec:formalism},
the formalism for optical phonon mediated drag in multiple subband
systems is presented.  In Section \ref{sec:calculation} the details
of the calculation and the results are presented, both for the case
of a single and double subband system.  Finally, the conclusions
are presented in Section \ref{sec:conclude}.

\section{Formalism}
\label{sec:formalism}

In this section, the generalization for the expression 
for the drag transresistivity for the case of several subbands is given. 
Then, the explicit form for polar optical phonon mediated interaction 
is introduced, following the formalism for phonon-mediated drag 
given by Zhang and Takahashi\cite{zhan93} and
B{\o}nsager et al.\cite{bons97}  

\subsection{Transresistivity with multiple subband systems}

In a typical drag experiment, two electrically isolated 2DEGs are
placed close together.  A current in passed through one layer
(denoted layer 1), which induces a voltage in the second (layer 2).
The quantity experimentalists measure is the transresistivity 
$\tensor{\rho}_{21}$, which is the ratio of the electric field 
in the second layer to the current density in the first layer
\begin{equation}
\tensor{\rho}_{21}\cdot {\bf J}_1 = {\bf E}_2.
\end{equation}
Naturally, the stronger the interlayer coupling,
the larger the magnitude of $\rho_{21}$.

We define the plane of the 2DEGs as the $xy$ plane.
We use upper case letters for 3-dimensional vectors and lower
case for vectors in the plane of the 2DEG; e.g., 
${\bf Q} = ({\bf q},Q_z)$.
The expression for the linear response transresistivity for multiple 
subbands systems,
in the semi-classical weak interlayer interaction approximation,
comes about from a simple generalization of the derivation in 
Ref.\ \onlinecite{flen95b}.
At temperature $T$, with density $n_i$ and charge $e_i$
of layer $i$ 
\begin{eqnarray}
\tensor\rho_{21}&=& 
-\frac{\hbar^2}{2\pi e_1e_2 n_1 n_2 k_B T}
\int \frac{d{\bf q}}{(2\pi)^2} \int_0^{\infty} d\omega\ 
\sum_{\alpha\beta\gamma\delta} 
\frac{{\bf F}_1(\alpha\beta;{\bf q},\omega)\; 
{\bf F}_2(\gamma\delta;{\bf q},\omega)\ 
|v_{12}(\alpha\beta,\gamma\delta;{\bf q},\omega)|^2}
{\sinh^2(\hbar\omega/2k_B T)}.
\label{rho_21}
\end{eqnarray}
Here $v_{12}(\alpha\beta\,\gamma\delta;{\bf q},\omega)$ is the
effective interlayer interaction which scatters carriers from subband 
$\alpha\rightarrow\beta$ in layer 1 and $\gamma\rightarrow\delta$ in 
layer 2, with the change of in-plane momentum $\hbar{\bf q}$
and energy $\hbar\omega$ (see Fig.\ \ref{schematic}). 
The functions ${\bf F}$ are given by
\begin{eqnarray}
{\bf F}(\alpha\beta;q,\omega) &=& \frac{2\pi e}{\hbar\mu_t}
\int \frac{d{\bf k}}{(2\pi)^2} 
[f_{0\alpha}({\bf k})-f_{0\beta}({\bf k}+{\bf q})]
\ \delta(\varepsilon_\alpha({\bf k}) - \varepsilon_\beta({\bf k + q})
-\hbar\omega) \times
\nonumber\\
&&\ \ \ \ \ \ \ ({\bf v}_\alpha({\bf k+q}) \tau_\alpha({\bf k+q}) 
- {\bf v}_\beta({\bf k})  \tau_\beta({\bf k}))
\end{eqnarray}
where $\mu_t$ is the mobility of the sample, ${\bf v}_\alpha$ is the
velocity, $\tau_\alpha$ is the relaxation time and $f_{0\alpha}$ 
is the Fermi-Dirac distribution function in subband $\alpha$.
This relaxation time is defined by the linear response 
$\delta\! f$ to a small electric field ${\bf E}$,
\begin{equation}
\delta\! f_\alpha({\bf k}) = 
\left(-\frac{\partial f_{0\alpha}}{\partial\varepsilon}\right)
e_i {\bf E}\cdot{\bf v}_\alpha({\bf k})\;\tau_\alpha({\bf k}).
\end{equation}

For parabolic isotropic systems where intralayer electron--electron
scattering dominates over other scattering mechanisms, the distribution 
functions in all subbands are drifted Fermi-Diracs.\cite{flen95b,hu96}
Then, the expression for the transresistivity becomes
\begin{eqnarray}
\rho_{21} = -\frac{h}{e^2}\frac{1}{4\pi} 
\int_0^\infty \frac{dq\ {q}^3}{k_F^4}
\int_0^\infty \frac{\hbar\ d\omega}{k_B T}
\sum_{\alpha\beta\gamma\delta}
\frac{
\chi_1''(\alpha\beta;q,\omega)\; 
\chi_2''(\gamma\delta;q,\omega)\ 
|v_{12}(\alpha\beta,\gamma\delta;q,\omega)|^2}
{\sinh^2(\hbar\omega/2 k_B T)}
\label{rho_chi}
\end{eqnarray}
where $\chi_i''$ is the imaginary part of
the random phase approximation irreducible polarizability
\begin{equation}
\chi_i''(\alpha\beta;q,\omega) = 
\int \frac{d{\bf k}}{2\pi} 
\left[f_{0\alpha}({\bf k}) - f_{0\beta}({\bf k + q})\right] 
\ \delta\left(\varepsilon_\alpha({\bf k}) - \varepsilon_\beta({\bf k + q})
-\hbar\omega\right)
\end{equation}
We shall assume that this form in valid for the remainder of the paper. 

\subsection{Interlayer phonon mediated interaction}

We assume that the properties of the phonon are the same
in both the well and the barrier region.  We denote the (bulk) electron-phonon 
interaction matrix element $M_\lambda({\bf q},Q_z)$, where $\lambda$ is the
polarization (for clarity the $\lambda$ will henceforth be suppressed), 

Let $\psi_{i\alpha}(z)$ be the transverse 
wavefunctions of the electrons in well $i$ and subband $\alpha$.
A generalization of the Ref. \onlinecite{zhan93} and \onlinecite{bons97}
to multiple subbands for an effective bare phonon propagator  
which scatters electrons from $\alpha\rightarrow\beta$ in well $i$ and 
$\gamma\rightarrow\delta$ in well $j$ gives 
\begin{equation}
d_{ij}^{(0)}(\alpha\beta,\gamma\delta;{\bf q},\omega) 
= \int_{-\infty}^{\infty} 
\frac{dQ_z}{2\pi}\ D^{(0)}({\bf q},Q_z,\omega)
|M({\bf q},q_{z})|^2\, f_i(\alpha\beta,Q_z) 
f_j(\gamma\delta,-Q_z),
\label{d_ij}
\end{equation}
where $f_i(\alpha\beta,Q_z)$ is the form factor given by
\begin{equation}
f_i(\alpha\beta,Q_z) = 
\int_{-\infty}^\infty\ dz\ \psi_{i\alpha}^*(z) \psi_{i\beta}(z) 
e^{-iQ_z z}
\end{equation}
and $D^{(0)}({\bf Q},\omega)$ is the bare (retarded) bulk phonon Green function.
This effective phonon propagator mediates the interaction between
the electrons in the adjacent wells.

\subsection{Longitudinal polar optical phonons}

The bulk longitudinal polar optic phonon--electron interaction matrix
element is given by\cite{mahan} 
\begin{equation}
|M^2(Q)| = V_c(Q)\  
\frac{\hbar\omega_{\rm LO}}{2}
\left(\frac{1}{\epsilon_\infty}-\frac{1}{\epsilon_0}\right)
\label{mq}
\end{equation}
where
\begin{equation}
V_c(Q) = \frac{4\pi e^2}{\epsilon_\infty Q^2}
\label{vc}
\end{equation}
is the bare Coulomb interaction.

We assume that the optical phonons have a constant 
frequency $\omega_{\rm LO}$, since the longitudinal optical phonons have a 
small dispersion at the Brillouin zone center. 
Also assuming a finite lifetime $\gamma_{\rm LO}^{-1}$ gives the form for the 
bare phonon Green function
\begin{equation}
D^{(0)}_{\rm LO}(\omega) =
\frac{1}{\omega - \omega_{\rm LO} + i\gamma_{\rm LO}/2}
- \frac{1}{\omega + \omega_{\rm LO} + i\gamma_{\rm LO}/2}.
\end{equation}
With these assumptions, the phonon Green function $D$ can be factored
out of Eq.\ (\ref{d_ij}), and the spatial form of the 
effective longitudinal polar optical phonon mediated interaction
is exactly the same as the Coulomb interaction.
Together with Eqs.\ (\ref{mq}) and (\ref{vc}) this gives 
\begin{eqnarray}
d_{ij,{\rm LO}}(\alpha\beta;\gamma\delta;{\bf q},\omega) 
&=&
D_{\rm LO}^{(0)}(\omega)\; \frac{\omega_{\rm LO}}{2} \;
\left(1 - \frac{\epsilon_\infty}{\epsilon_0}\right)
V_{ij}^c(\alpha\beta,\gamma\delta; q)
\label{d_ret}
\end{eqnarray}
where $V_{ij}^c(\alpha\beta,\gamma\delta; q)$ is the 
Coulomb matrix element scattering between $\alpha\rightarrow\beta$ 
in well $i$ and $\gamma\rightarrow\delta$ in well $j$, together
with an in-plane change of $\hbar{\bf q}$.

\subsection{Effective screened interlayer interaction}

The optical phonon contribution to the drag rate is obtained to lowest
order by substituting a $d_{12,{\rm LO}}^{(0)}$ into 
$v_{12}$ in Eq. (\ref{rho_21}).\cite{interfere}
It has been shown, however, screening of the interaction gives 
many novel effects in drag experiments.\cite{flen95b,flen94,bons96}
In principle screening within the random phase approximation can be 
included for systems where two subbands participate in each layer, 
but one needs to invert an $16\times 16$ matrix to find the screened
interaction.
However, we shall see that screening is not an important 
consideration in the case of optical phonon mediated drag.

One important effect of screening is to remove 
a spurious mathematical divergence in the transresistivity
in Eq. (\ref{rho_21}).  If the phonon lifetime is infinite 
($\gamma  = 0$) then in the absence of screening the bare phonon 
interaction gives a nonintegrable divergent momentum transfer at the
phonon frequency $\omega_{\rm ph}(q)$ for all $q$.  Even if the lifetime 
is finite but large, the screening of the electron--phonon interaction is
the dominant physical mechanism for elimination of the
divergence in the integrand around $\omega_{\rm ph}(q)$.\cite{bons97}
In the case of the acoustic phonons, the lifetime is very large 
(on the order of a microsecond) because acoustic phonons are unable 
to decay via anharmonic terms to other phonons, and hence proper
treatment of the screening is important to obtain a quantitatively
correct result.  On the other hand, the optical phonons
can decay via anharmonic terms, leading to much shorter intrinsic lifetimes 
(on the order of picoseconds).  In this case, the broadening of the phonon
Green function due to the lifetime is the physically
dominant mechanism for the removal of the divergence around
$\omega_{\rm ph}(q)$.  

Furthermore, as shown below, most of contribution to the 
integral Eq. (\ref{rho_21}) comes from the small $q$ and 
$\omega\approx \omega_{\rm LO}$ regime, where the effects of screening 
are small.
Therefore, given the short lifetime of the longitudinal polar
optical phonons and the weakness of the screening in the $\omega-q$
space important for optical phonons, it is justified
to use the bare interaction $d^{(0)}$ as the interlayer 
interaction.\cite{intersubplas}  

\section{Calculation and results}
\label{sec:calculation}

Longitudinal optical phonons are well defined excitations 
when $\gamma_{\rm LO}$ is much smaller than $\omega_{\rm LO}$.
Since $\gamma_{\rm LO}$ is also much smaller than all other 
frequency scales in the problem, one can approximate  
\begin{equation}
|D_{\rm LO}^{(0)}(\omega)|^2 
\approx \frac{2\pi}{\gamma_{\rm LO}} \left[\delta(\omega - \omega_{\rm LO})
+ \delta(\omega + \omega_{\rm LO})\right].
\label{deltafn_approx}
\end{equation}
Note that the real (imaginary) parts of $D_{\rm LO}^{(0)}(\omega)$ 
corresponds to the virtual (real) phonon contribution.\cite{bons97}
In the case of optical phonon mediated interaction, the virtual and 
real phonons contribute equally to the drag.

Using Eq.\ (\ref{deltafn_approx}) and Eq.\ (\ref{d_ret})
in Eq.\ (\ref{rho_chi}) and the other approximations stated above, 
the polar optical phonon mediated drag transresistivity is given by 
\begin{eqnarray}
\rho_{21} 
&\approx& 
\frac{1}{8}\;\frac{h}{e^2}\; 
\left(1 - \frac{\epsilon_\infty}{\epsilon_0}\right)^2 
\frac{\omega_{\rm LO}}{\gamma_{\rm LO}}\,
\frac{\hbar\omega_{\rm LO}}{k_B T} 
\frac{1}{\sinh^2(\hbar\omega_{\rm LO}/2k_B T)}
\nonumber\\ 
&&\ \ \ \ \ \ \int_0^\infty d\tilde q\ \tilde{q}^3 
\sum_{\alpha\beta\gamma\delta}
\chi_1''(\alpha\beta;\tilde{q},\omega_{\rm LO})
\chi_2''(\gamma\delta;\tilde{q},\omega_{\rm LO})
\,\left| V_{12}^c(\alpha\beta,\gamma\delta;\tilde{q})\right|^2
\label{rho21}
\end{eqnarray}
where $\tilde{q} = q/k_F$.
Note that the apparent divergence of the above result 
$\gamma_{\rm LO}\rightarrow 0$ is an artifact of the neglect of
the dynamical screening scheme, as discussed in the previous
section.  

We now discuss drag in the case when one has single and double
subbands in each layer.

\subsection{Drag in single-subband systems}

In the case of drag when there is only a single subband occupied,
the expression Eq. (\ref{rho_chi}) reduces to the well-known
one given previously.\cite{jauh93,zhen93}
The main difference between previous treatments and 
optical phonon mediated drag is that optical phonon scattering 
involves a relatively large exchange of energy,   
which necessarily implies that the in-plane transfer of momentum is
quite large.

In the case when an electron emits or absorbs an optical phonon, 
its kinetic energy in the direction of the plane changes by 
$\hbar(\omega_{LO}-\omega_{\alpha\beta})$, where $\omega_{\alpha\beta}$
is the energy change between the subbands $\alpha$ and $\beta$. 
The momentum transfer $\delta{\bf q}$ associated with this for a
parabolic band is 
\begin{equation}
\delta{\bf q}\cdot \langle {\bf v}\rangle = \omega_{LO}-\omega_{\alpha\beta},
\label{delta_q}
\end{equation}
where $\langle{\bf v}\rangle$ is the velocity of the average
of the initial and final state.  In the case where the electron stays
within the same subband, $\omega_{\alpha\beta} = 0$, and hence 
$q \gtrsim \omega_{LO}/\langle v\rangle$.

We insert numbers for GaAs.  For density $n =  1.5\times 10^{11}\,
{\rm cm}^{-2}$,
$\langle v \rangle \approx v_F \approx 1.7 \times 10^7 {\rm cm/s}$, 
and $\omega_{LO} \approx 6 \times 10^{13}\ {\rm s}^{-1}$, one obtains a
typical $q$ of approximately $3 \times 10^6 \,{\rm cm}^{-1}$.
Given that the typical distance separating the wells $d$ is on the
order of $400\,{\rm\AA}$, giving a product $q d \approx 14$. 
Since the Coulomb interaction (and hence by Eq.\ (\ref{d_ret}) the 
optical phonon mediated interaction) goes as $\exp(-q d)$, and
the scattering strength is given by the matrix element squared,
one obtains a factor of $e^{-28}$, which makes it almost impossible to
see the effect.

In order to see an effect in single subband systems it is clear
that one must reduce the typical momentum transfer $\delta{\bf q}$.
To do this, one must have a lower optical phonon frequency  
and a larger $\langle v\rangle$; i.e., a smaller effective mass.
A possible candidate is InSb.
It has an effective mass of approximately $0.02 m_e$ where $m_e$ is the
bare electron mass, and a longitudinal
optic phonon frequency of $24\,$meV (compared to $0.067 m_e$ and
$36\,$meV, respectively, for GaAs).  
Assuming an electron density $n = 1.5 \times 10^{11}\,{\rm cm}^{-2}$, 
$d = 300\,{\rm\AA}$ and well-width 
$L = 200\,{\rm\AA}$, the integral in Eq. (\ref{rho21}) for 
$k_B T\lesssim \hbar\omega_{\rm LO}$ is 
approximately $1.9 \times 10^{-3}$. 
The coupling constant
$(1 - \epsilon_\infty/\epsilon_0)^2 = 5\times 10^{-3}$,
yielding 
\begin{equation}
\rho_{21} \approx 30\,{\rm m}\Omega \ \frac{\hbar\omega_{\rm LO}}{k_B T}  
\frac{1}{\sinh^2(\hbar\omega_{\rm LO}/2k_B T)} \ 
\frac{\omega_{\rm LO}}{\gamma_{\rm LO}}
\label{rho21-1subband}
\end{equation}
For $k_B T = \hbar\omega_{\rm LO}/2$, the temperature terms are
equal to 1.44.  Therefore, we have for the parameters mentioned above 
\begin{equation}
\rho_{\rm InAs}(T = 110\,{\rm K}) \approx 40\,{\rm m}\Omega\,
\frac{\omega_{\rm LO}}{\gamma_{\rm LO}}
\end{equation}

The result depends on the product of the optical phonon frequency
and lifetime.  In a relatively clean system, this ratio is large.
Using GaAs as a guide, the lifetime $\tau\approx 7\,{\rm ps}$,
at 77 K.\cite{phon_life}
Therefore, $\gamma_{\rm LO} = 6.6 \times 10^{-16} {\rm eV}/5 \times 10^{-12}
\approx 0.1\,{\rm meV}$, giving  $\omega_{\rm LO}/\gamma_{\rm LO} \approx 360$.
This leaves one with a transresistivity of on the order of $1 - 10\,\Omega$,
which should be experimentally observable. 

\subsection{Double subband systems}

We define the lower subband as $A$ and upper subband as $B$. 
In double-subband systems, one can tune $\omega_{BA}$ so that 
it is equal to $\omega_{\rm LO}$.  
Then, from Eq.\ (\ref{delta_q}) the minimum momentum 
transfer is zero, and therefore there can be significant contribution
from small in-plane momentum transfer events, without the cut-off in 
the exponential  
in $V^c_{21}(q)$.  As a consequence, we shall see that the drag rate can 
be rather large. 

We assume the wells are made of GaAs.  As mentioned above, the intrasubband
excitations do not give significant contributions.  Therefore
the main contributions come from the terms $\chi_i''(AB;q,\omega)$.
Since the integral in Eq. (\ref{rho_chi}) is restricted to $\omega > 0$, 
the only term is the sum over subband indices which contributes significantly
for $k_B T \lesssim \hbar\omega_{\rm LO}$ is $\chi_1(AB;q,\omega) 
\chi_2(AB;q,\omega).$
In this case, using a square well potential we have 
\begin{eqnarray}
V_{12}^c(AB,AB;q) &=& 
\left[\frac{16 \cosh(qL/2) \pi^2 qL}  
{9\pi^2 + 10\pi^2 (qL)^2 + (qL)^4}\right]^2
\end{eqnarray}

Using the above in Eq. (\ref{rho21}) allows us to calculate $\rho_{21}$.
For GaAs, the term $h/(8e^2) (1 - \epsilon_\infty/\epsilon_0)^2 \approx
65\,\Omega$.  
For temperatures $k_B T \lesssim \hbar\omega_{\rm LO}$ at typical
experimental densities, the
term in the integrand in Eq. (\ref{rho21}) is not a strong function
of temperature and hence can be evaluated in the $T=0$ limit (when 
$k_B T \gtrsim \hbar\omega_{\rm LO}$ the lowest subband becomes
nondegenerate and a more careful evaluation of the temperature 
dependence of $\chi_i''$ in Eq.\ (\ref{rho21}) is necessary).
Fig.\ \ref{well_width} shows this integral in Eq.\ (\ref{rho21}) 
as a function of the energy difference $\omega_{AB}$.
The drag rate peaks as $\omega_{AB}=\omega_{\rm LO}$ for the reasons
stated above.  Estimating the ratio $\omega_{\rm LO}/\gamma_{\rm LO}
\approx 200$,\cite{phon_life} we see that the at $T = 100\,{\rm K}$,
the optical phonon mediated drag is on the order of $30\,\Omega$.
As the subband separation moves further away from $\omega_{\rm LO}$,
the transresistivity decreases rapidly. 
The optical phonon mediated interaction is unique in that 
it is particularly extremely sensitive to width of the wells.

\section{Conclusions}
\label{sec:conclude}

We have calculated the transresistivity for coupled quantum wells
due the mediation of polar optical phonons.  
The spatial form of the polar optical phonon is identical to that
of the Coulomb interaction.  We find that in GaAs, when only a 
single subband is important (i.e., $\hbar\omega_{AB}\gg E_F, k_B T$), 
the transresistivity is negligible at present experimental conditions. 
This is because the polar optical phonon scattering is an inelastic
event resulting in a minimum $q$-vector which substantially reduces
the effect.  However, it may be possible to see single subband
drag in other materials with lower $\omega_{\rm LO}$ and lower effective
masses such as InSb. 
In the case where there are two different subbands, the
subband energy difference can be engineered to be the same as
the $\omega_{LO}$, in which case the optical phonon mediated 
transresistivity can be relatively large and observable.  
The magnitude of the drag should
be sensitive to the subband energy separation relative to the 
optical phonon frequencies.  It may be possible to observe this
resonance by tuning the subband energy splitting through 
$\omega_{\rm LO}$ by means of an external gate which changes
the shape of the confining potential in the wells.

In this work we have assumed that the
intralayer electron--electron interactions are strong enough
that the effective $\tau$ is constant (i.e., a drifted Fermi Dirac
distribution for all subbands).  For the clean samples used in
these experiments, this is generally the case.
We have also assumed that the 
phonon frequencies in the barrier and in the well are identical. 
In fact the optical phonons in ${\rm Al}_x{\rm Ga}_{1-x}$As 
have frequencies which are, depending of the $x$, somewhat higher than
GaAs.  The peak shown in Fig. \ref{well_width} would tend to be
smeared by this mismatch in frequencies.

The author thanks M. C. B{\o}nsager, K. Flensberg, T. J. Gramila, 
A.-P. Jauho and A. H. MacDonald for useful discussions and comments.

\begin{figure}
\caption{
Schematic for the double quantum wells system with two subbands,
denoted $A$ for lower and $B$ for upper.  Experimentally 
current is passed through layer 1 and a voltage is measured in
layer 2.   The phonon-mediated scattering is indicated by the 
Feynman diagram.  The filled dots are the electron--phonon vertices 
and the dotted line indicates the phonon Green function.
The $\alpha\beta\gamma\delta$ are the subband indices, ${\bf k},{\bf
k'}$ are the in-plane momenta before collision and ${\bf q}$ is 
the in-plane momentum transfer.  
}
\label{schematic}
\end{figure}

\begin{figure}
\caption{
The integral in Eq.\ (\protect\ref{rho21}), 
$d = 400\,{\rm\AA}$, $n = 1.5 \times 10^{11}\,{\rm cm}^{-2}$,
and $\omega_{\rm LO} = 36\,{\rm meV} \approx 400\,{\rm K}$, as in GaAs.   
We assume infinite square wells for the confinement, in which case 
$\omega_{AB} = \omega_{\rm LO}$ for $L \approx 215$\,\AA.
Inset: temperature dependence of the transresistivity for low
temperatures, $x/\sinh^2(1/2x)$ as a function of $x=\hbar\omega_{\rm
LO}/k_B T$ [see Eqs.\ (\ref{rho21}) and (\ref{rho21-1subband})].
}
\label{well_width}
\end{figure}


\begin{thebibliography}{10}
\bibitem{interfere}
In principle, the bare interlayer interaction $v_{12}$ is the sum of the
  Coulomb term $V^c_{12}$ and phonon term $d$, and there are interference terms
  between these two terms. However, the optical phonon contribution is strong
  and sharp and localized in $q-\omega$ space, and hence the interference
  effects $V^{c}_{12} d$ are minimal compared to the ``purely phonon'' term
  $d^2$.
\bibitem{intersubplas}
One possible screening effect which might complicate the situation is coupling
  the optical phonons to intersubband plasmons (A. Kuznetsov, private
  communication). As the intersubband plasmons always lie above the
  intersubband single-particle continuum due to the depolarization shift, they
  do not directly participate in the drag, but they can couple to the optical
  phonons and shift the frequency of that mode. The main qualitative effects
  mentioned in this paper, however, should not be altered.
\end{thebibliography}
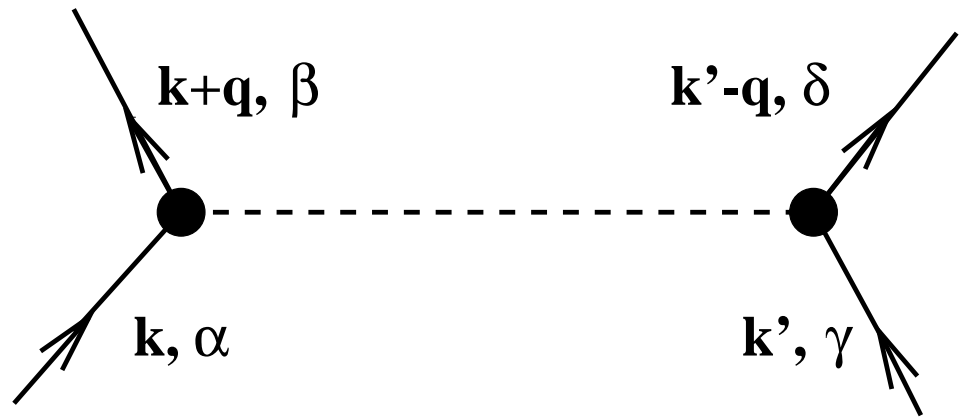
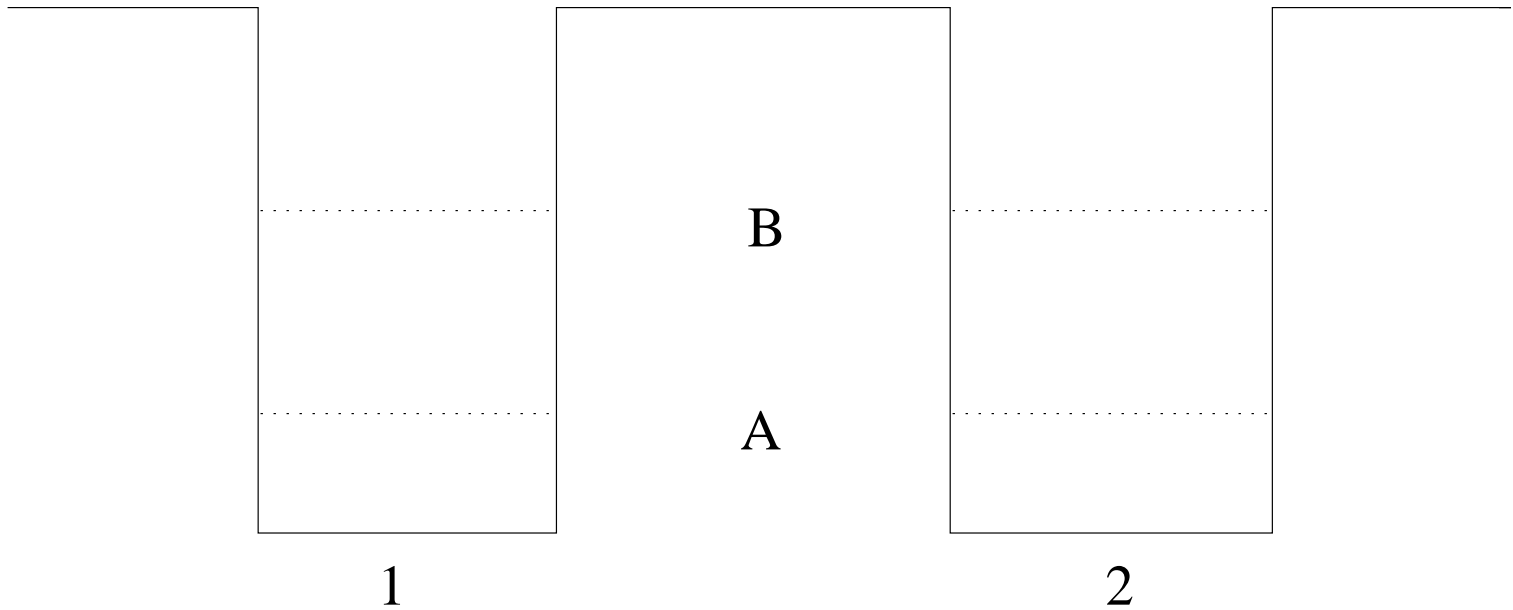
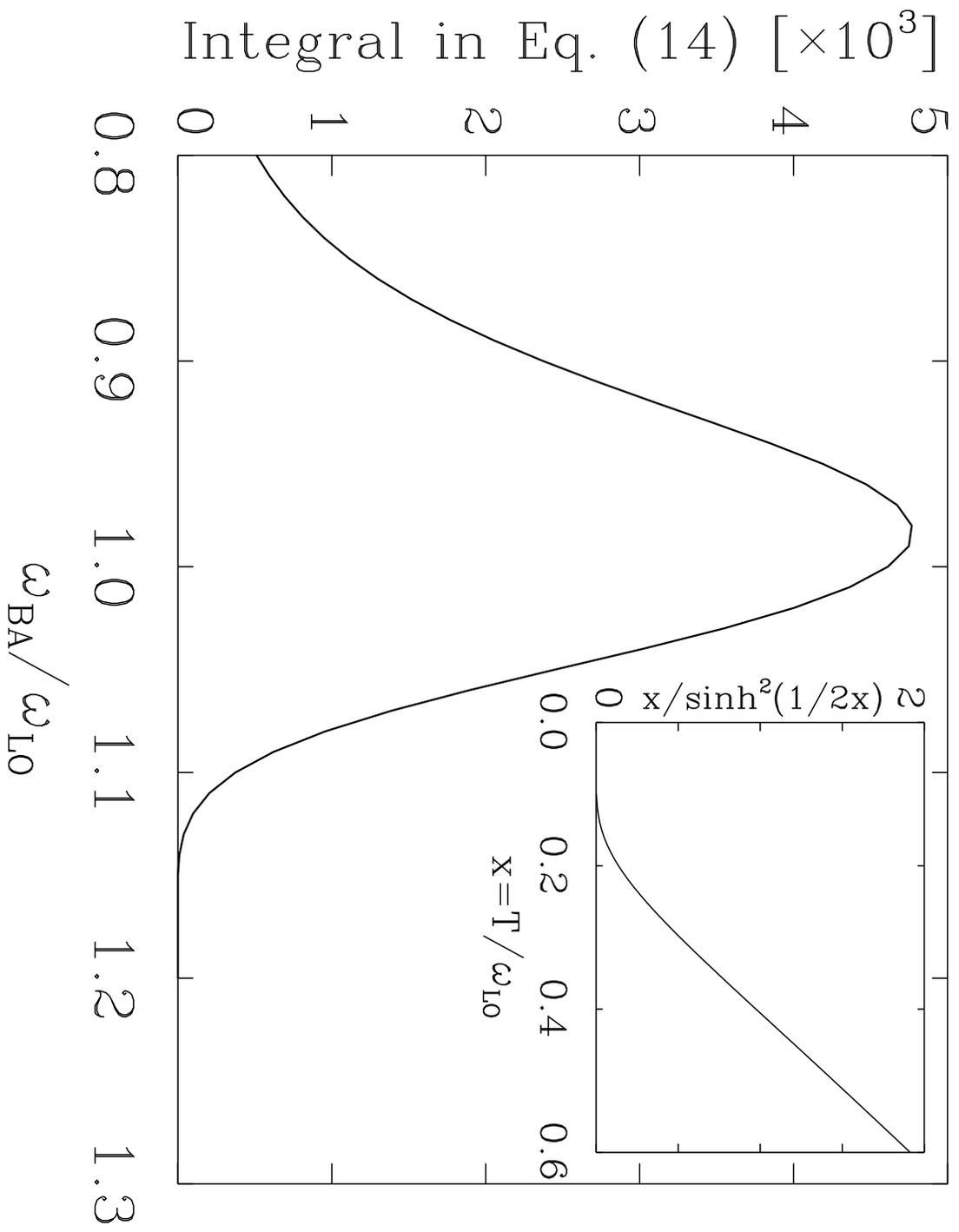
\end{document}